# The Labor Economics of Paid Crowdsourcing

John J. Horton
Harvard University
383 Pforzheimer Mail Center
56 Linnaean Street
Cambridge, MA 02138
john.joseph.horton@gmail.com

Lydia B. Chilton
University of Washington
AC101 Paul G. Allen Center, Box 352350
185 Stevens Way
Seattle, WA 98195-2350
hmslydia@cs.washington.edu

## ABSTRACT

Crowdsourcing is a form of "peer production" in which work traditionally performed by an employee is outsourced to an "undefined, generally large group of people in the form of an open call." We present a model of workers supplying labor to paid crowdsourcing projects. We also introduce a novel method for estimating a worker's *reservation wage*— the smallest wage a worker is willing to accept for a task and the key parameter in our labor supply model. It shows that the reservation wages of a sample of workers from Amazon's Mechanical Turk (AMT) are approximately log normally distributed, with a median wage of $1.38/hour. At the median wage, the point elasticity of extensive labor supply is 0.43. We discuss how to use our calibrated model to make predictions in applied work. Two experimental tests of the model show that many workers respond rationally to offered incentives. However, a non-trivial fraction of subjects appear to set earnings targets. These "target earners" consider not just the offered wage—which is what the rational model predicts—but also their proximity to earnings goals. Interestingly, a number of workers clearly prefer earning total amounts *evenly divisible by 5*, presumably because these amounts make good targets.

## Categories and Subject Descriptors

J.4 [**Social and Behavioral Sciences**]: Economics;; J.m [**Computer Applications**]: Miscellaneous

## 1. INTRODUCTION

Crowdsourcing is a form of "peer production" that outsources work traditionally performed by an employee to an "undefined, generally large group of people in the form of an open call" [2, 11]. Despite the successes and the perceived promise of crowdsourcing,[1] would-be users face a serious practical challenge: they need to attract a crowd. Crowdsourcing projects have used a variety of inducements, including entertainment [20], information [1, 13], the chance to be altruistic and attention from others [12]. Until recently, it was extremely difficult to offer a crowd money, but with the advent of online labor markets like oDesk, Elance and Amazon's Mechanical Turk (AMT), buyers can now easily pay workers with cash [8].

Compared to cash, non-monetary crowdsourcing incentives are limited in at least two ways. First, some non-monetary incentives depend on the nature of the task or the identity of the proposer, which limits their usefulness. For example, tasks such as classifying web pages, transcribing scanned documents and validating search results are ideal for crowdsourcing, yet they are unlikely to attract volunteers because they are tedious and the private benefits come only to the proposer. Second, even when appropriate non-monetary incentives exist, they are hard to adjust. For example, assuming one could compute a "fun" labor supply elasticity, $\epsilon$, it is not clear how we could make a task $10(1/\epsilon)\%$ more fun in order to boost provision by 10%. With cash incentives, computing supply elasticities is straightforward and making price adjustments is easy.

The possibility of cash payments raises a design question: how does one effectively and efficiently employ monetary incentives? To solve this problem, designers need two things: (1) a theoretical model that predicts how people will respond to different price/task scenarios and (2) data-driven refinements of that model. The refinements should allow the model to account for behavioral biases—such as those identified by experimental economics—and the idiosyncrasies of particular applications. Borrowing an analogy from the economist and market designer Al Roth, just as builders of suspension bridges must be informed by both the elegant theory of mechanics and the messy details of metallurgy and geology, builders of social systems must possess both a general theory of behavior and detailed contextual knowledge [17].

An appropriate theoretical model for the labor supply aspect of crowdsourcing design should tell the designer (1) how workers decide whether or not to participate in a crowdsourcing project and (2) how workers decide the amount to produce, conditional upon participating. In the language of economics, the designer must be able to predict the labor supply on both the (1) extensive and (2) intensive margins. These two aspects of labor supply have intrigued economists from Adam Smith onward, but because crowdsourcing "jobs" last seconds and pay pennies, results from conventional labor economics might not transfer readily (or at least not without modification). This caveat aside, labor economics offers a theoretical framework for understanding decision-making in paid crowdsourcing scenarios, and it potentially provides the predictive model that designers need.

---

[1] Examples include Wikipedia, Digg, Yelp, Yahoo! Answers and InnoCentive.

## 1.1 Overview

There are several papers on the design of incentives in crowdsourcing [5] and on obtaining work from AMT [18, 19], but our paper is most closely related to work by Watts and Mason [16], who also conducted labor supply experiments. They found that workers respond to prices in a way that is at least *consistent* with rational behavior. However, their findings also suggest that the relationship between incentives and output is complex: higher pay rates did not improve work quality—a result the authors attribute to high wages affecting worker beliefs. They also found that the structure of incentives (i.e., whether workers faced a piece rate or a quota system) affected output.

Motivated by the evidence from [16] that AMT workers respond to cash incentives, in Section 2.1 we develop a simple rational model of crowdsourcing labor supply. We also present a novel method for estimating the reservation wage of each worker (Section 2.3) that makes use of a highly concave earnings function rather than a simple piece-rate.[2] The reservation wage is the minimum wage a worker is willing to accept as compensation in exchange for performing some task; it is the key parameter in models of labor supply.

With subjects recruited from AMT, we test the predictions of the model and its underlying assumption in two separate experiments. Subjects performed simple tasks in which they chose how much output to produce. We test whether subjects adjusted output in response to task difficulty (Experiment A, Section 4) and price (Experiment B, Section 5). We find mixed evidence for the rational model: workers are clearly sensitive to price but insensitive to variations in the amount of time it takes to complete a task.

Reservation wages are supposed to be fixed and thus should be invariant to our experimental manipulations. Although we find that the imputed wage distributions are quite similar in Experiment A, large differences appear in Experiment B. The cause seems to be that some workers workers are "target earners" who focus on reaching salient earnings targets. This stands in sharp contrast to the rational model that predicts workers should only consider the offered wage. These findings have important implications for the design of incentives and are discussed in Section 6.

The rational model clearly misses important elements of reality, but there are no immediate replacements and it makes reasonable predictions in some cases. For these reasons, we demonstrate in Section 7 how the model can be used to predict labor supply for any price/task scenario, using calibration data pooled from both experiments. We conclude with a discussion of the contributions and limitations of the paper and our thoughts on directions for future research.

## 2. THEORY

Every time-consuming activity generates an *opportunity cost*. The opportunity cost of doing A is the foregone net benefits one would have obtained from doing a next best option B. Researchers generally cannot observe the net benefits of a person's options, but when they observe them doing A, they can infer that they find doing A preferable to doing B, with all rewards and costs for the two tasks being taken into account. Applying this inference to work decisions yields a prediction: a person will work only when the net benefits from working exceed the hypothetical net benefits from their next best alternative, be it another job, leisure or a renewed job search. In labor models, the economic value of this "next best alternative" is characterized as a reservation wage.

The reservation wage is difficult to estimate in practice for at least two reasons. First, all jobs offer a mixture of non-monetary benefits and costs, or amenities and dis-amenities. For example, there are obvious non-monetary differences between working as a coal miner and working as an ice cream taste-tester. Second, observing someone working tells us only that their total benefits exceed total costs, but not what those total cost actually are. Imagine a job offers a wage $w$ and a stream of amenities, $a$, and a stream of dis-amenities, $d$. If the worker works for time $t$, they receive benefits $(w + a)t$ and bear costs $dt$. If the worker has a reservation wage, $\omega$, then observing someone working tells us only that $w + a - d \geq \omega$.

To estimate $\omega$, we need to identify the worker's indifference point, i.e., the $w^*$ where $w^* + a - d = \omega$. To push a worker down to his or her indifference point, we could continuously lower their wage by small amounts until they chose to quit. Assuming workers viewed this process sanguinely and their marginal costs were not increasing, their wage when they exit—i.e., when they are indifferent between working and continuing—is their reservation wage for that task. This "decreasing wage" method is clearly impractical in traditional labor relationships, but [4] show that it can easily be done for small, piece-rate tasks when the process is explained up front and workers have little emotional investment in their seconds-old "job." In our experiments, we capitalize on this feature of piece-rate work and and use a continuously decreasing payment function.

## 2.1 Model of labor supply

Because the payments in crowdsourcing contexts are so small, we do not expect a worker's marginal utility of wealth to change while working and thus we assume $u(w) = w$. Workers choose some positive, continuous quantity to produce, $y \geq 0$. They are paid $P(y)$ for their choice of $y$. Let $P'(y) = p(y)$. We assume that $P(y)$ is strictly increasing, $p() > 0$, and concave, $p'() < 0$. We assume that it costs the worker $C(y)$ to complete $y$ tasks, with $C'(y) = c(y)$. The

---
[2]All of our data and experimental materials will be available on John Horton's webpage: http://www.people.fas.harvard.edu/~horton.

worker's maximization problem is:

$$\max_y P(y) - C(y) \quad s.t. \quad y \geq 0 \quad (1)$$

The first-order condition is $p(y^*) = c(y^*)$, i.e., the marginal benefit equals marginal cost. An interior solution exists only when $B(y) = P(y) - C(y)$ reaches a global maximum, which occurs when $B(y)$ is concave.

## 2.2 Cost curves and output

In most applied contexts, $P()$ will be linear (i.e., a constant piece-rate), so $B(y)$ will be concave only when $-C(y)$ is strictly concave, i.e., marginal costs are increasing. If a task is very tiring, marginal costs are increasing ($c'(y) > 0$), but this is not necessarily the case: costs could be decreasing if workers get better with experience ($c'(y) < 0$), and costs can also be approximately linear ($c'(y) = 0$) or even have different properties at different points (e.g., decreasing at first, then linear).

Consistent with our opportunity cost framework, we assume that the only costs of performing a task is the time it takes: $C(y) = \int_0^y \omega t(x) dx$ where $t()$ is the marginal completion time and hence $c(y) = t(y)\omega$. The advantage of this assumption is that because the $t(y)$ *curve* is observable, we can tell whether costs are increasing, decreasing or constant. In our experiments, we find that completion times are essentially constant and we assume linear costs.[3] This linearity is unsurprising given the simplicity and brevity of our tasks.

In practice $P()$ is often linear, with a constant piece-rate $\pi$ for each unit of output, giving a payment function $P(y) = \pi y$. If costs are linear, there is no interior solution and workers are willing to produce either nothing when $\omega > \pi/t$ or an infinite amount when $\omega < \pi/t$. The actual manifestation of this pheneomena is that worker's either produce nothing or produce all the way up to some cap set by employers. For example, in the experiments run by [16], workers could choose how many picture sorting tasks to perform, up to a cap of 100 and with each task paying a constant amount. In the experimenter's high-wage, low-difficulty condition, *mean* output was over 90 tasks, and presumably many of the subjects hit the 100 task cap.

## 2.3 Using the payment function to impute the reservation wage

With linear costs and a linear payment function (constant piece-rate), an interior solution to the maximization problem does not exist. However, if we make $P(y)$ strictly concave by having the piece-rate fall with greater output, then an interior solution at $p(y^*) = \omega t_0$ exists. A worker's estimated reservation wage is then estimated directly from their output choice: if a worker completes $y_i^*$, then $\hat{\omega}_i = \frac{p(y_i^*)}{\bar{t}_i}$, where $\bar{t}_i$ is the worker's average completion time. If costs were increasing, then instead of a point estimate of costs, $\hat{t}_0$, we use some parametric prediction, $\hat{t}_i(y_i^*)$ (recall that we observe completion times).

---
[3]Excluding the very first task, completion times rise by less than 1 *millisecond* per task. There is, however, a gap between the first task and second task, with the second task requiring about 1.5 fewer seconds

For ease of exposition, we modeled worker output as continuous. In most practical crowdsourcing applications, subjects make discrete output choices. Subjects chose some number of whole number of tasks to complet and $P(y) = \int_0^y p(x) dx$. The discrete analog to this function is $P(y) = \sum_{x=0}^y p(x)$, hence $p(y+1) = P(y+1) - P(y)$. When output is discrete, measuring reservation wages is somewhat more complex, but we know that $\omega_i \leq p(y_i^*)/\bar{t}_i$ and $\omega_i > p(y_i^*+1)/\bar{t}_i$. If the piece-rate tasks are small and output is fine-grained, then:

$$\hat{\omega}_i \approx (2\bar{t}_i)^{-1} [p(y_i^*) + p(y_i^* + 1)]$$

reasonably approximates the reservation wage.

To use this method, it is important that workers are aware that the marginal payment is falling and will continue to fall. Otherwise, incorrect expectations that wages might increase eventually might cause them to continue working even when wages fall below their reservation wage.

## 3. EXPERIMENTAL PRELIMINARIES

Any reasonable labor supply model should predict that lowering wages will reduce output. There are two ways to lower a person's wages: (a) increase the output they must produce in order to earn their previous wage or (b) lower their wage while keeping the required amount of work constant. As we will show, our model predicts that output will fall in either scenario. We then test these predictions in two experiments. In Experiment A, we test whther increasing the task difficulty reduces output, and in Experiment B, we test whether lowering wages reduces output. Both of our experiments had the same basic set-up: workers from AMT were asked to perform piece-rate tasks. Critically, they had complete freedom to choose how many pieces to complete. Before we discuss the experiments in depth, we first provide necessary background information.

## 3.1 Amazon's Mechanical Turk

Amazon's Mechanical Turk is an online labor market where workers can perform "Human Intelligence Tasks" (HIT) for "requesters." HITs vary, but most are small, simple tasks that are difficult for computers, but relatively easy for humans. Common tasks include transcribing audio clips, classifying and tagging images, reviewing documents and checking websites for pornographic content. When posting a HIT, a requester describes the task, sets a piece-rate payment, sets worker qualifications, determines how long workers can work on the task, determines how many times they want each HIT performed and creates an interface for workers to use when working on the task.

To become an AMT worker, a person must create an AMT account and provide Amazon with a bank account number. Workers are only allowed to have one account, and Amazon uses several technical and legal means to enforce this restriction. Workers can observe the collection of HITs and their attributes prior to starting work and can normally view a sample of the required work before starting. They are free to work on any task for which they are qualified, and they can begin work immediately after accepting a HIT. Once a worker completes a HIT, they must submit it for review. A requester then may review the work and decide whether or not to "approve" the HIT. If the HIT is approved, the worker is paid the piece-rate. The worker is also paid if the requester does not review and approve the work within a specified amount of time. Solely at their discretion, requesters may "reject" work, in which case the worker is not paid.[4] Requesters may also elect to pay bonuses, which makes it easy to tailor payments to individual workers based on their performance within a nominally piece-rate HIT.

## 3.2 Conduct of the experiments

Experiment A and B were conducted in sequence, with approximately 3 days between them. Of the 92 subjects who participated in A, 38 of them also participated in B, which had 198 subjects.

Subjects were not informed that the task had an experimental component. The HIT was simply posted on AMT like any other HIT, with the task title "User interface test." It is important to note that all subjects read identical instructions before accepting the HIT and before observing any unique feature of their assigned treatment group. Thus we are confident that subjects did not select out of the experiment in response to the nature of their assigned treatment. In both experiments, all subjects accepting the HIT submitted a response, so no outcome data are missing.[5]

## 3.3 Task and user interface

For a crowdsourcing task, we had subjects click back and forth between two narrow vertical bars separated by some number of pixels. The "target" bar was green and the other bar was gray. The target bar alternated between the left bar and the right bar, with the first target bar always begining on the left. If they missed a bar, the bar flashed red but workers could continue. Figure 1 represents the interface and the cursor movement needed to complete the task.

We chose the clicking task because it is time consuming, requires the full attention of subjects and is not obviously fun. This task is also culturally neutral, easy to understand and easy to make harder (by narrowing the bars or spacing the bars farther apart). We organized work into *blocks*, with each block consisting of 10 back-and-forth clicks. A block is one unit of output (i.e., if a worker completes 3 blocks, $y = 3$). Subjects had 4 seconds to complete each *click*. If they did not perform the click within this time, the HIT ended, but across both experiments, no subject exceeded this cutoff. If a subject missed more than 40% of the clicks in a block, the HIT ended. As with the time cutoff, no subject came anywhere close to this limit. We capped total output at 200 blocks, but this proved unnecessary as the observed maximum output was only 58 blocks.

After completing a block, subjects chose whether to quit or continue. If they had already completed $y - 1$ blocks, they were offered $p(y)$ to perform an additional block, where, $p(y) = P(y) - P(y-1)$. If they chose to quit, they performed no more tasks and earned $P(y - 1)$. When making this exit decision, the interface showed then all the relevant rates and totals, as well as their error rate and their average per-block completion time in seconds. Subjects could rest as long as they liked before starting a new block.

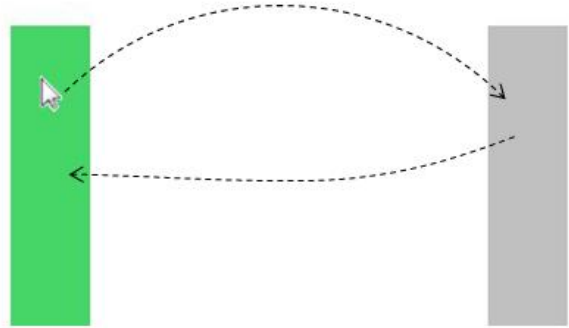

Figure 1: **Crowdsourcing task. Subjects were asked to click between the two vertical rectangles. The "target" bar (shown on left here) was always colored green.**

## 3.4 Payment function for our experiments

To create a concave payment function, each subsequent piece of work must pay less than the previous piece of work, but still offer positive payment. One payment function with this property is:

$$P(y) = \bar{P}\left(1 - e^{-ky}\right) \qquad (2)$$

Note that total earnings asymptotically approach $\bar{P}$ as $y$ increases. The parameter $k \geq 0$ can be thought of as determining the "half-life" of the payment schedule. For example, if a given $k^*$ has the property that $P(10, k^*) = \frac{1}{2}\bar{P}$, then the half-life of the schedule is 10 tasks. In all experiments, we used a half-life of 10; Table 1 shows samples of the total earnings and the marginal payment using Equation 2, given this half-life.[6]

Table 1: **Sample earnings and wages with 10 task half-life**

| $y$ | $P(y)/\bar{P}$ | $p(y+1)/\bar{P}$ |
|---|---|---|
| 1 | 0.07 | 0.0625 |
| 5 | 0.29 | 0.0474 |
| 25 | 0.82 | 0.0118 |

One complication in our payment scheme is that it is impossible to pay workers fractional cents. To circumvent this problem, we used a payment system where a worker was paid all of their whole-cent earnings for sure, and their fractional earnings stochastically. If a worker earned $h$ whole cents and $f$ fractions of a cent, with probability $1 - f$, we paid them $h$ and with probability $f$, we paid them $h + 1$. Payment was thus correct in expectation, since $\mathbb{E}[P(y)] = (1 - f)h + (h + 1)f = h + f$. This procedure was explained to subjects before they joined the experiment.

---

[4][10] provide a model of the accept/reject decision in markets like AMT and show under what conditions an "all reject" equilibrium does not occur.

[5]A discussion of the causal inference issues raised by revealed-preference experiments conducted online can be found in [9].

[6]For subjects producing only one unit of output, an adjustment must be made when imputing reservation wages beacuse $P(1)$ needs to incorporate the "show-up" fee.

## 4. EXPERIMENT A: Δ DIFFICULTY

In Experiment A, subjects were randomly assigned to groups $EASY$ and $HARD$. The only difference between the two groups was that in $EASY$ the vertical bars were 100 pixels apart compared to 600 pixels apart in $HARD$. Of the 92 subjects that participated, 42 were assigned to $EASY$ and 38 were self-reported females. Subjects completed a total of 18934 clicks. In both groups, subjects' earnings were determined using the same payment function, Equation 2, with the parameter values of $\check{P} = 10$ and $k = 10^{-1} \log 1/2$. With these parameter values, total earnings asymptotically approached 10 cents and the "half-life" was 10 blocks, i.e., $P(10) = 5$.

### 4.1 Model prediction

Using the first-order condition from Equation 1, $p(y) = \omega t$ (for simplicity, we drop the $*$ notation) and treating optimal output as a function of $t$, we take the total derivative with respect to $t$ and get $y'(t) = \omega/p'(y(t))$. Because $p'() < 0$, it follows that $y'(t) < 0$. As expected, increasing the unit completion time reduces a worker's output.

### 4.2 Results

#### 4.2.1 Effort

Usurprisingly, subjects assigned to $HARD$ needed more time to complete a block. Regressing average per block completion time (in seconds), $\bar{T}_i$, on the treatment indicator $EASY_i$ (with robust standard errors under each coefficient) [7] we have:

$$\bar{T}_i = \underbrace{-4.885}_{1.25} \cdot EASY_i + \underbrace{10.931}_{1.2}$$

with $R^2 = 0.13$ and sample size $N = 92$. The treatment effect is large and highly significant: subjects in $HARD$ took about 11 seconds to complete a block; subjects in $EASY$ only needed about 6 seconds. If we examine between-click times instead of average block completion times, we see further evidence that the treatment was effective. In Figure 2, the distributions for both "hits" an "misses" are shown for each group. As expected, the "hit" distribution for $HARD$ is right-shifted. Confirming our intuitions about the relationship between effort and quality, misses were associated with faster cursor movement.

#### 4.2.2 Output

Even though subjects in $HARD$ needed more time to complete each block, they had nearly the same pattern of output as subjects in $EASY$. Figure 3 shows output histograms for both groups. There is no discernible difference in mean output, and although although more workers in $HARD$ quit after performing just one block (12 vs. 7), this difference is not statistically significant.[8] Regressing output on a group indicator we have:

$$y_i = \underbrace{-0.247}_{3.76} \cdot EASY_i + \underbrace{20.08}_{2.72}$$

with $R^2 = 5e - 05$ and the sample size 92. Subjects in $EASY$ had slightly less average output, though the effect

---
[7] All standard errors in the paper are robust.
[8] We regressed $1\{y_i = 1\}$ on $EASY_i$.

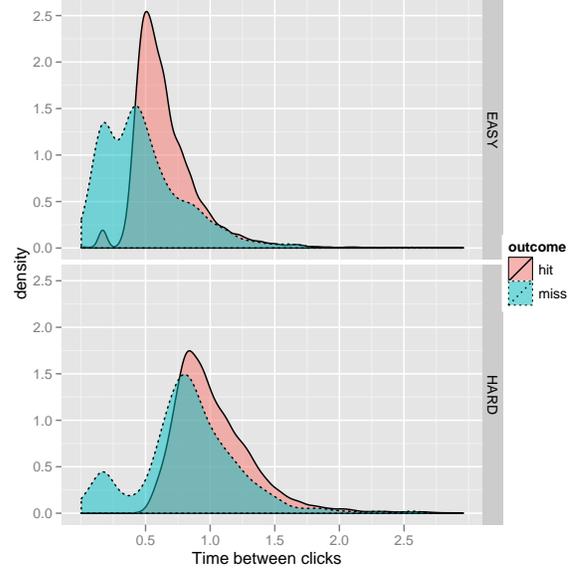

**Figure 2:** Distribution of time-between-clicks for "hits" and "misses" by treatment group. Distributions are computed using a kernel density estimator.

is imprecisely estimated. Transforming output with the log function and running the same regression we have:

$$\log y_i = \underbrace{0.136}_{0.28} \cdot EASY_i + \underbrace{2.298}_{0.2}$$

with $R^2 = 0.00256$ and $N = 92$. Even though the coefficient on $EASY$ changes signs, both the regressions and the graphical analysis of Figure 3 lead to the same conclusion: despite the greater time required per-block for subjects in $HARD$, there was no discernible difference in the patterns of output across groups.

#### 4.2.3 Reservation wages

We imputed the reservation wage for each subject using the method explained in Section 2.3. The distribution of the log of estimates is plotted in Figure 4, which shows both the smooth kernel density estimate of the distributions as well as the the actual estimated values (displayed as tick marks along the horizontal axis). The top two panels contain the results of Experiment A. We can see that the imputed reservation wage distributions are quite similar, except that $HARD$ has a fat left tail, implying that a cluster of workers in $HARD$ exhibited output patterns consistent with very low reservation wages. However, this could be due to sampling variance. Indeed, in a regression of log reservation wages on a group indicator we have:

$$\log \hat{\omega}_i = \underbrace{0.523}_{0.37} \cdot EASY_i + \underbrace{-0.117}_{0.26}$$

with $R^2 = 0.02$ and sample size $N = 92$. Assignment to $EASY$ had a positive but statistically insignificant effect. The coefficients in this regression are more easily interpretable following transformation. They imply that the geometric mean reservation wage was $0.89/hour in *HARD* and $1.5/hour in *EASY*.

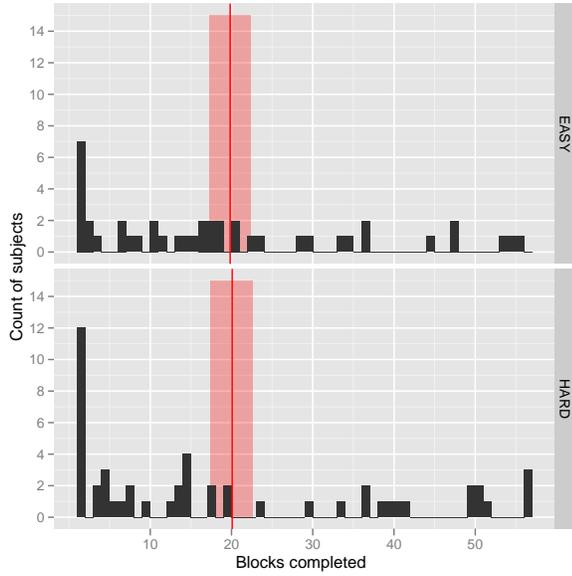

**Figure 3: Output by task difficulty.** The red vertical line indicates mean output in both groups, while the shaded band shows plus/minus the standard error in the estimated mean. Bins have unit length.

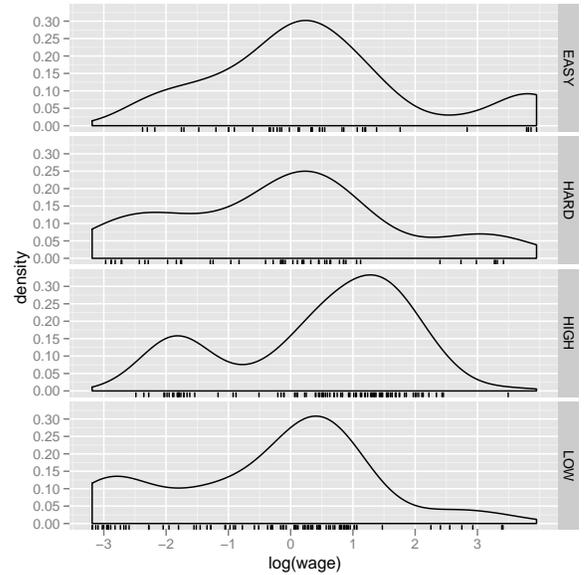

**Figure 4: Imputed log reservation wage distributions both experimental groups, in both experiments.**

### 4.3 Discussion
The experimental results are theoretically ambiguous but are internally consistent. Given that workers in *HARD* took longer per task but had essentially the same output pattern compared to subjects in *EASY*, we would expect subjects in *HARD* to have relatively higher reservation wages, which is what we find, albeit the effect only has a t-statistic of 1.41.

Perhaps the simplest explanation for the lack of treatment effects is that our treatment was not strong enough and that workers are not attuned to fairly small differences in time. The quote by Lord Lionel Robbins about the "marginal utility of not bothering with marignal utility" seems particular apt, especially if worker's have a heuristic task-based (as opposed to time based) reservation rate.

## 5. EXPERIMENT B: ∆ PRICE
In Experiment B, 198 subjects were randomly assigned to groups *HIGH* and *LOW*. The task was identical in both groups, with the bars 100 pixels apart, but earnings asymptotically approached either 10 cents (in *LOW*) or 30 cents (in *HIGH*). Because of the imprecise treatment effect estimates in Experiment A, we doubled the sample size. By chance, both groups had the same number of subjects. Of the 198 subjects, 72 were self-reported females. Subjects completed a total of 45710 clicks.

### 5.1 Model prediction
Consider an alternative payment function $\gamma P(y)$, where $\gamma$ is some positive number. The first order condition for the original maximization problem is now $\gamma p(y) - t\omega = 0$, and if we treat the optimal output $y$ as a function of $\gamma$, the total derivative of the first order condition with respect to $\gamma$ is $p(y) + \gamma p'(y(\gamma))y'(\gamma) = 0$ and thus:

$$y'(\gamma) = -\frac{p(y)}{\gamma p'(y(\gamma))}$$

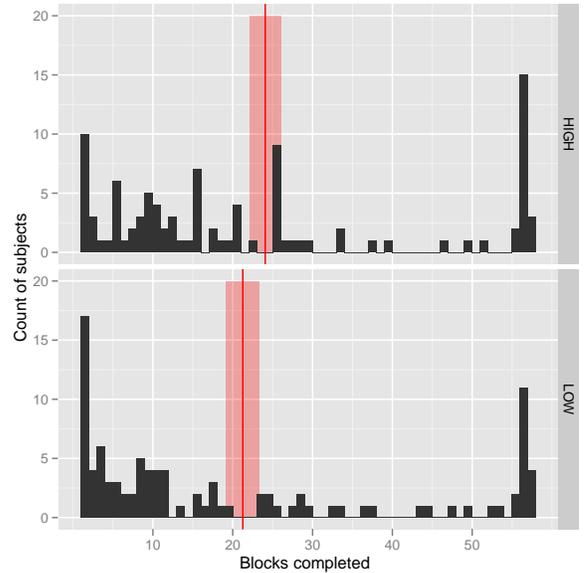

**Figure 5: Output by price group.** The red vertical line indicates mean output in both groups, while the shaded band shows plus/minus the standard error in the estimated mean. Bins have unit length.

Because payment is strictly increasing, $p() > 0$, and because the total payment function is cocave, $p'() < 0$, it follows that $y'(\gamma) > 0$, i.e., output increases when payment is higher.

## 5.2 Results

### 5.2.1 Output

Figure 5 shows the histogram of output in the two groups. Although $LOW$ has more early quits, both groups have sizable numbers of early quitters as well as some stalwarts that completed in excess of 50 blocks. Unlike in Experiment A, mean output in $LOW$ is noticeably lower than in $HIGH$. Regressing output on a group indicator we have:

$$y_i = \underbrace{-2.808}_{2.89} \cdot LOW_i + \underbrace{24.071}_{2}$$

with $R^2 = 0.0048$ and $N = 198$. Subjects in $LOW$ had lower mean output, but the effect is imprecisely estimated due to the bi-modality of output, which can be seen in the output histograms. Estimating the regression in logs rather than levels we have:

$$\log y_i = \underbrace{-0.303}_{0.17} \cdot LOW_i + \underbrace{2.712}_{0.11}$$

with $R^2 = 0.02$ and $N = 198$. There is a large and significant difference in geometric means across the groups. From Figure 5, it appears that most of the effect is due to the relatively large number of subjects in $LOW$ quitting after small amounts of output. We can confirm this graphical observation by regressing an indicator for whether a subject completed fewer than 10 blocks on the treatment indicator:

$$1\{y_i < 10\} = \underbrace{0.152}_{0.07} \cdot LOW_i + \underbrace{0.273}_{0.05}$$

with $R^2 = 0.03$ and $N = 198$. The choice of 10 blocks is somewhat arbitrary, but the effect is strong for all low cutoff values, and a QQ-plot (not shown) makes the pattern even more apparent.

### 5.2.2 Reservation Wages

Unlike in Experiment A, group assignment strongly affected the imputed reservation wage. The bottom two panels of Figure 4 show that the mass of observations at the center of the distribution is left-shifted for $LOW$ relative to $HIGH$. Both distributions also have a second mass of observations with very low reservation wages, but as with the main hump, the $LOW$ observersations are shifted further left than the low-wage hump in $HIGH$. Regressing log wages on a group indicator we have:

$$\log \hat{\omega}_i = \underbrace{-0.792}_{0.22} \cdot LOW_i + \underbrace{0.447}_{0.14}$$

with $R^2 = 0.06$ and $N = 198$. Transforming the predictions into levels, the geometric mean reservation wage was $1.56/hour in $HIGH$ and $0.71/hour in $LOW$.

## 5.3 Discussion

As per the model's prediction, lower pay reduces output, but the finding that being in $LOW$ lowers a worker's reservation wage implies that the lower output in $LOW$ is not as low as it *should* be. Because the actual tasks are identical across groups, the difference in imputed reservation wages cannot be due to uncontrolled-for differences in amenities. There are several other possible explanations for the data, including worker error, increasing but non-time based marginal costs and target earning behavior; we believe target earning provides the most compelling explanation for our findings.

Worker error could explain our results: we would find lower reservation wages in $LOW$ if (1) subjects in $LOW$ falsely believe that they are being paid more than they actually are, (2) subjects in $HIGH$ falsely believe that they are being paid less than they actually are or (3) some combination of (1) and (2). Liebman and Zeckhauser [15] argue that people systematically misinterpret schedules, usually by letting early marginal payments "bleed over" to affect perceptions of future marginal payments. However, with our concave schedules, this bleed over would occur in both groups. While we have no evidence on this question, it seems likely that any bleed over effects would be greater in $HIGH$, which would lead to effects in the opposite direction of what we do find. Furthermore, because subjects were informed of their precise marginal payment after each block, this bleed over explanation seems unlikely.

Constant marginal costs are a key assuption in our reservation wage estimation method. If marginal costs are rapidly increasing, then we would incorrectly infer that subjects in $LOW$ had reservation wages too low: the additional output subjects in $HIGH$ should produce because of their higher wages is moderated by the higher costs of high output. Although an increasing marginal cost explanation is possible, it seems unlikley, given that subjects could rest for as much time as they liked between blocks. Furthermore, even subjects producing lots of output spent less than 15 minutes total, including resting time (recall that per block time for the 100 pixel group was less than 7 seconds). Rapid increases in marginal costs over such a short duration of time seem rather unlikely.[9]

An additional explanation for the results is that some workers may be *target earners*. Target earners try to obtain some self-imposed earnings goal rather than respond to the current offered wage. If at least some workers are target earners, then the results are easy to explain: because their wages are lower, subjects in $LOW$ must produce more output to acheive their earnings targets than they would have to produce if they were instead assigned to $HIGH$.

## 6. DEPARTURES FROM THE RATIONAL MODEL

The differences in the imputed reservation wage distribution—particularly those found in Experiment B—strongly suggest that the rational model cannot explain the behavior of all workers. Our best conjecture is that workers create earnings targets that influence their output decisions. While having goals seems sensible and perhaps heuristically "rational" if the target provides motivation, earnings targets lead workers to commit a kind of sunk cost fallacy because, in the absence of income effects, past earnings are irrelevant to the decision they must make at the margin (i.e., whether the next bit of earnings is worth the the trouble of the next bit of work).

---

[9]One possibility we must consider with more complex work is that workers are likely to need more experience with a task before deciding whether they are good at it. In these cases, we might mistakenly view learning and exit as increasing marginal costs.

To return to the suspension bridge building analogy from Section 1, we are now dealing with metallurgy instead of mechanics: target earners do not behave like rational workers in certain contexts, and this difference will matter in applications. For example, in the absence of income effects, when wages are high, a target earner works less and a rational worker works more. There is mixed evidence on the question of whether target earning occurs in "real life" [3, 6, 7], but we find fairly unambiguous evidence of target earning in several places in the results.

The first piece of evidence of target earning is that in every experiment, at least some subjects try to pursue the maximum earnings possible, despite the low wages associated with this strategy. For rational workers to generate this pattern, the wage distribution would have to be highly bi-modal. A more plausible explanation is that workers try to earn the full amount possible (i.e., $\bar{P}$ is their target) and only quit when they realize this goal is unattainable.[10]

### 6.1 Preferences for "focal point" earnings

More evidence of target earning can be see in the pattern of output. In Figure 6, we plot histograms of the output for *HIGH*, grouped in horizontal panels by the floor of earnings, $\lfloor P(y) \rfloor$. Subjects show a preference for working the minimum amount possible to earn some *whole* number of cents: when multiple output values yield the same number of whole cents, the far left bar of the histogram is the highest. The only exception is in the 29 cent panel, however, recall that in *HIGH* subjects' earnings asymptotically approached 30 cents. Subjects quit once they realized that they could not break out of the 29 cent band. Although suggestive of targeting, the preference for whole cents could be symptomatic of extreme risk aversion (recall from Section 3.2 that payment is stochastic due to the fractional cent problem), or workers could believe that we might cheat them on the fractional cents (i.e., ignore their fractional earnings) but that we would not be willing to withhold whole-cent earnings.

Harder to reconcile with a rational model is another pattern seen in Figure 6: subjects show a preference for earnings amounts evenly divisible by 5. The smallest earnings amounts (e.g., 2, 3 and 5 cents) all get several subjects, but because these are people who quit very early, they presumably do not have a target or would not need a target. However, we see clear output spikes at 15, 20 and 25 cents.

---

[10] A more pedestrian explanation could be that some subjects believe that the payment is in dollars, not cents, despite the clear instructions that stated that payment would be in cents. One subject did email us insisting that payment was supposed to be in dollars—we sent him screen shots proving this was not the case.

Assume that in the absence of "modulo 5" targeting, the proportion of subjects earning amounts divisible by 5 should be equal, in expectation, to the proportion of potentially realizable amounts divisible by 5. Consider the set of possible whole-cent earnings:

$$\mathbf{P_w} = \{\lfloor P(1) \rfloor, \lfloor P(2) \rfloor, \ldots \lfloor P(N) \rfloor\}$$

and the fraction of those elements of $\mathbf{P_w}$ that are divisible by 5, which is given by:

$$q = |\mathbf{P_w}|^{-1} \sum_{x \in \mathbf{P_w}} 1\{x \mod 5 = 0\}$$

where $|.|$ indicates the number of elements in the set. Let $\mathbf{Y}$ be the set of realized output choices for an experiment group and let $n$ be the number of observations, $n = |\mathbf{Y}|$ and let $s$ be the actual number of observations in $Y$ such that the whole-cent earnings were divisible by 5: $s = \sum_{y \in \mathbf{Y}} 1\{\lfloor P(y) \rfloor \mod 5 = 0\}$

Under our assumption of proportionally, we can compute the probability of observing $s$ successes out of $n$ trials when the per-trial probability of success is $q$. We had 33 successes out of 99 trials when the probability was only $q = 0.22$. The probability of observing this many successes or more by chance is only 0.0027.

## 7. CONCLUSION

We find some agreement with a simple rational model, as well as important anomalies; we find fairly strong evidence that at least some worker's work to targets. Designers should consider this propensity when designing incentives schemes and give people natural targets that will increase output, though they should also consider that such schemes might seem manipulative and could backfire (and potentially be unethical).

In the following section, we demonstrate how our calibrated model can be used in applied work. While we only find partial agreement between the model's predictions and reality, we beleive it still offers a useful approximation. We conclude by discussing how our paper fits into a larger research program and lay out some directions for future investigations..

### 7.1 Using the calibrated model

It is straightforward to use our calibrated model for prediction. Consider some crowdsourcing task that takes $t_0$ to complete and that will pay a piece-rate of $p_0$. Workers whose reservation wage exceeds the offered wage, $p_0/t_0$, will accept the task. The fraction of workers producing at least one unit of output is equal to the probability that a given worker will find participating attractive: $Pr(p_0 \geq \omega_i t_0) = F\left(\frac{p_0}{t_0}\right)$ where $F$ is the cumulative density function of the estimated reservation wage distribution.

The labor supply curve is $S(w) = N_s F(\log w)$, where $N_s$ is the number of workers in the population and $w = p_0/t_0$. The point elasticity of extensive labor supply is thus $\epsilon_w = f(\log w)/F(\log w)$. To compute the intensive elasticity, we would have to know something about the change in marginal costs.[11]

---

[11] If we learn that $c(y) = \omega y t + (y-1)^2 \nu$, output on the

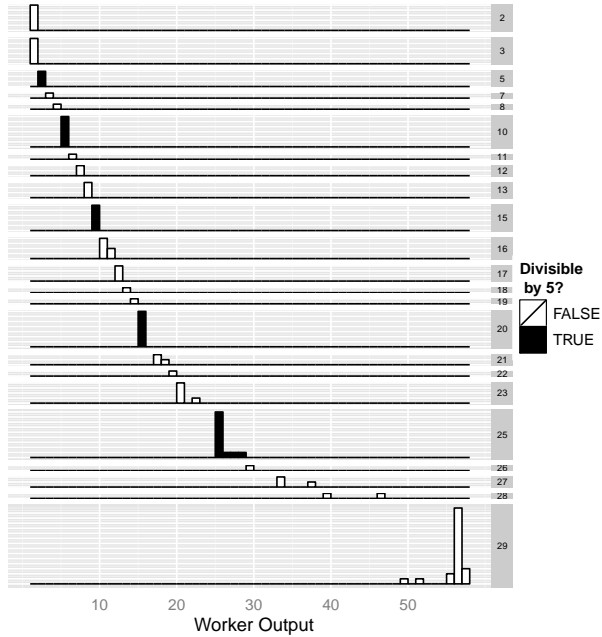

**Figure 6: Earnings in the *HIGH* Group.** This plot shows the distribution of output, grouped by the whole digit of worker earnings (in cents). The vertical axis scale is not shown and is scaled differently by groups. However, each horizontal white line indicates a gap of one.

If we pool estimates from both experiments and assume that wages are log normally distributed, the distribution parameters are $\hat{\mu} = 0.074$ and $\hat{\sigma} = 1.634$, with the reservation wage measured in dollars per hour. In Table 2, the 25th, 50th and 75th percentiles of the pooled reservation wage distribution are shown, as well as the extensive labor supply elasticity computed at that point using the log normal approximation. Consisent with a right-skewed distribution like the log normal, the arithmetic mean is considerably higher than the median wage.

**Table 2: Pooled reservation wage distribution properties**

|  | Wage ($/hour) | Point Elasticity |
|---|---|---|
| 25th p. | 0.321 | 0.81 |
| Median | 1.384 | 0.43 |
| 75th p. | 2.876 | 0.28 |
| Mean | 3.625 | 0.24 |

It is important to remember the fairly strong assumptions underlying these predictions. First, the reservation wage distribution was estimated using a selected sample of AMT workers willing to participate in our experiment. Second, we are assuming constant marginal costs, and we are assuming that task amenities/dis-amenities for the new task are equivalent to those from the back-and-forth clicking task we used. Third, we are assuming that there is a fixed reserva-

---

intensive margin can be predicted: each worker produces $y^*$, such that $p(y^*) - \omega t + 2\nu y^* = 0$.

tion wage and that workers respond rationally to the offered wage, despite some evidence to the contrary.

## 7.2 Future research

Crowdsourcing is still a new development and many open questions remain. Perhaps the most obvious next empirical step is to collect data on the compensating differentials associated with different kinds of crowdsourcing tasks. In other words, compared to a baseline task, how much more or less do workers have to be paid to generate the same amount of output? A related question is how costs change with output, i.e., does a task get much easier or much harder as the worker gains experience? Finally, it would be interesting to learn about the correlates of reservation wages. For example, do groups with lower opportunity costs (e.g., the unemployed, non-US citizens, etc.) have lower reservation wages?

Perhaps the biggest limitation of the current model is that it only predicts the fraction of workers that will accept a task. It tells us nothing about how many workers will actually see a given offer on AMT. In our experience, the time since a HIT was posted affects uptake, presumably because "fresh" HITs are easier to find when searching by date posted. Furthermore, AMT workers are not uniformly distributed across time zones and presumably the number of potential workers waxes and wanes over the day. It would be useful to learn more about search and uptake and it might shed light on the process of job search—a topic of great interest to economists.

## 7.3 Main contribution

We view our paper as a step towards the theoretical framework for crowdsourcing/human computation advocated by [14]. Jain and Parkes argue (correctly in our view) that advances in crowdsourcing as a methodology will require broadly applicable, predictive models. They discuss game theory as a potential generalizing framework, but point out that crowdsourcing "games" are not so much about workers revealing private information (which would suggest a mechanism design approach) as they are about getting workers to show up and exert the effort needed to accomplish a task.

In a labor relationship, there is a "game" between workers and their employers: workers must choose how much (costly) effort to provide and employers—who prefer high effort—cannot directly observe this choice. This classic principal/agent problem arises in many real-world contexts, but we argue that in *most* crowdsourcing applications, it is a fairly easy problem to solve.[12] When effort is highly correlated with output and output is observable, simple mechanisms can eliminate moral hazard. By rejecting low quality work (or not hiring low quality workers in the future), buyers can easily make shirking a dominated strategy.

Although the moral hazard problem is solvable, the problem of getting a sufficient supply of labor persists. Once an employer sets up a quality control mechanism (e.g., screening, firing, spot checks, etc.), the worker is essentially playing a game against nature. Individual workers will make labor supply decisions by comparing the costs and benefits of working, and although workers must think rationally about their preferences, they do not have to think strategically. In short, we do not need a game theory of crowdsourcing, but rather a *price theory* of crowdsourcing. Our model and reservation wage estimation procedure provide the groundwork for such a theory.

## 8. ACKNOWLEDGMENTS

John Horton thanks the Berkman Center for Internet and Society and the NSF-IGERT Multidisciplinary Program in Inequality & Social Policy for generous financial support. Thanks to Richard Zeckhauser for extremely useful comments and suggestions. All plots were made using the open source R package `ggplot2`, developed by Hadley Wickham [21].

## 9. REFERENCES

[1] L. Adamic, J. Zhang, E. Bakshy, and M. Ackerman. Knowledge sharing and yahoo answers: everyone knows something. 2008.
[2] Y. Benkler. *The Wealth of Networks: How Social Production Transforms Markets and Freedom*. Yale University Press, 2007.
[3] C. Camerer, L. Babcock, G. Loewenstein, and R. Thaler. Labor supply of new york city cabdrivers: One day at a time. *The Quarterly Journal of Economics*, 112(2):407–441, 1997.
[4] D. L. Chen and J. J. Horton. The wages of paycuts: Evidence from a field experiment. *Working Paper*, 2009.
[5] D. DiPalantino and M. Vojnovic. Crowdsourcing and all-pay auctions. In *Proceedings of the tenth ACM conference on Electronic commerce*, pages 119–128. ACM, 2009.
[6] H. S. Farber. Reference-dependent preferences and labor supply: The case of new york city taxi drivers. *American Economic Review*, 98(3):1069—1082, 2008.
[7] E. Fehr and L. Goette. Do workers work more if wages are high? evidence from a randomized field experiment. *American Economic Review*, 97(1):298–317, 2007.
[8] B. Frei. Paid crowdsourcing: Current state & progress toward mainstream business use. *Produced by Smartsheet.com*, 2009.
[9] J. J. Horton and R. J. Zeckhauser. The potential of online experiments. *Working Paper*, 2010.
[10] J. J. Horton and X. Zhu. The allocation of decision rights in designed markets. *Working Paper*, 2010.
[11] J. Howe. *Crowdsourcing: Why the power of the crowd is driving the future of business*. Three Rivers Press, 2009.
[12] B. A. Huberman, D. Romero, and F. Wu. Crowdsourcing, attention and productivity. *Journal of Information Science (in press)*, 2009.
[13] S. Jain, Y. Chen, and D. Parkes. Designing incentives for online question and answer forums. In *Proceedings of the tenth ACM conference on Electronic commerce*, pages 129–138. ACM, 2009.
[14] S. Jain and D. C. Parkes. The role of game theory of human computation systems. In *HCOMP 2009*. ACM, 2009.
[15] J. Liebman and R. Zeckhauser. Schmeduling. *Working Paper*, 2004.
[16] W. A. Mason and D. J. Watts. Financial incentives and the 'performance of crowds'. *Knowledge Discovery and Data Mining- Human Computation (KDD-HCOMP)*, 2009.
[17] A. E. Roth. The economist as engineer: Game theory, experimentation, and computation as tools for design economics. *Econometrica*, 70:1341–1378, 2002.
[18] V. S. Sheng, F. Provost, and P. G. Ipeirotis. Get another label? improving data quality and data mining using multiple, noisy labelers. *Knowledge Discovery and Data Minding 2008 (KDD-2008)*, 2008.
[19] R. Snow, B. O'Connor, D. Jurafsky, and A. Y. Ng. Cheap and fast—but is it good? evaluating non-expert annotations for natural language tasks. *Proceedings of the Conference on Empirical Methods in Natural Language Processing (EMNLP 2008)*, 2008.
[20] L. Von Ahn. Games with a purpose. *IEEE Computer Magazine*, 39(6):96–98, 2006.
[21] H. Wickham. ggplot2: An implementation of the grammar of graphics. *R package version 0.7, URL: http://CRAN. R-project. org/package= ggplot2*, 2008.

---

[12] We qualify this statement because problems do arise when buyers do not know "what right looks like" and aggregating the inputs of other workers is not helpful for determining quality. For an example of this problem, see http://groups.csail.mit.edu/uid/deneme/?p=90.